\documentclass[twocolumn,amsmath,amssymb,aps,prl,longbibliography,floatfix]{revtex4-1}
\usepackage{physics}
\usepackage{units}
\usepackage{amsmath}
\usepackage{url}
\usepackage{natbib}
\usepackage{textcase}
\usepackage{amssymb}
\usepackage{graphicx}
\usepackage{bm} 

\usepackage{times}
\usepackage{float}
\usepackage{multirow,microtype,color,relsize,ulem}
\usepackage{subfigure}
\usepackage[breaklinks=true]{hyperref}
\usepackage[utf8]{inputenc}
\usepackage[english]{babel}
\hypersetup{colorlinks=true,linkcolor=blue,citecolor=blue}
\hypersetup{linktocpage}
\usepackage{CJK}
\usepackage{breakcites}
\usepackage{float}
\usepackage{siunitx}
\usepackage[dvipsnames]{xcolor}
\definecolor{mypine}{RGB}{1, 121, 111}

\newcommand{\req}[1]{Eq.\,(\ref{#1})}

\newcommand{\rfig}[1]{Figure\,\ref{#1}}

\newcommand{\rref}[1]{Ref.\,\onlinecite{#1}}

\def\tq{\tau_{\rm q}}

\begin{document}
\begin{CJK*}{UTF8}{gbsn}
\title{Isotropically conducting (hidden) quantum Hall stripe phases in a two-dimensional electron gas}

\author{Yi Huang~(黄奕)}
\email[Corresponding author: ]{huan1756@umn.edu}
\author{M. Sammon}
\author{M.\,A. Zudov}
\author{B.\,I. Shklovskii} 
\affiliation{School of Physics and Astronomy, University of Minnesota, Minneapolis, Minnesota 55455, USA}

\received{\today}

\begin{abstract}
Quantum Hall stripe (QHS) phases, predicted by the Hartree-Fock theory, are manifested in GaAs-based two-dimensional electron gases as giant resistance anisotropies.
Here, we predict a ``hidden'' QHS phase which exhibits \emph{isotropic} resistivity whose value, determined by the density of states of QHS, is independent of the Landau index $N$ and is inversely proportional to the Drude conductivity at zero magnetic field.  
At high enough $N$, this phase yields to an Ando-Unemura-Coleridge-Zawadski-Sachrajda phase in which the resistivity is proportional to $1/N$ and to the ratio of quantum and transport lifetimes. Experimental observation of this border should allow one to find the quantum relaxation time.
\end{abstract}

\date{\today}

\maketitle
\end{CJK*}

Quantum Hall stripe (QHS) phases in spin-resolved Landau levels (LLs) near half-integer filling factors $\nu = 9/2,11/2,13/2,...$, were predicted by the Hartree-Fock (HF) theory \citep{koulakov:1996,fogler:1996,moessner:1996}.
These phases consist of alternating stripes with filling factors $\nu \pm 1/2$, which, at exactly half-filling, both have the width $\Lambda/2 \simeq 1.42 R_c$ \citep{koulakov:1996,fogler:1996,wexler:2001,rezayi:1999}, where $R_c$ is the cyclotron radius. 
QHSs are formed due to a repulsive box-like interaction of electrons with ring-like wave functions. 
Such an unusual interaction leads to an energy gain when electrons occupy the nearest states within the same stripe and avoid interacting with electrons in neighboring stripes. 
The self-consistent HF theory is valid at LL indices $N \gg 1$, when $R_c =l_B(2N+1)^{1/2} \gg l_B$, where $l_B = (c\hbar/eB)^{1/2}$ is the magnetic length, a measure of quantum fluctuations of an electron's cyclotron orbit center, and $B$ is the magnetic field.
These fluctuations play a minor role even at $N=2$, and QHSs determine the ground state for all $\nu \ge 9/2$ \citep{fogler:1996,wexler:2001,rezayi:1999}.  

\begin{figure}[t]
\vspace{-0.2 in}
\includegraphics[width=\linewidth]{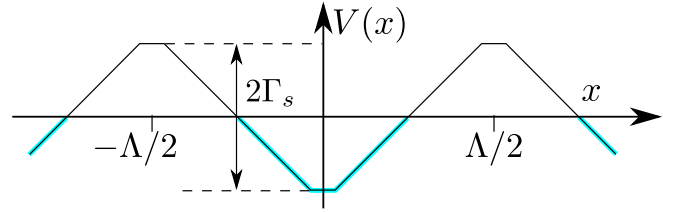}
\vspace{-0.2 in}
\caption{HF potential energy $V(x)$ responsible for QHS formation 
\citep{fogler:1996} at half-integer $\nu$.
The slope of $V(x)$ determines the internal electric field $\vb{E}$.
States shown by thick (cyan) lines are populated by electrons.
$\Lambda$ is the $V(x)$ period and $\Gamma_s$ is its amplitude.}
\label{fig:gamma_s}
\vspace{-0.1 in}
\end{figure}

QHSs were confirmed by the discovery of dramatic resistance anisotropies in two-dimensional electron gases in GaAs/AlGaAs heterostructures \citep{lilly:1999a,du:1999}. 
These anisotropies emerge because the diffusion mechanisms along and perpendicular to the stripe orientation are different \citep{macdonald:2000}.
In the stripe direction ($\hat y$) electrons drift along the stripe edge in the internal electric field $\vb{E}$ until they are scattered to an adjacent stripe edge by impurities.
If such scattering is weak, this mechanism leads to a large diffusion coefficient in the $\hat y$ direction (large conductivity $\sigma_{yy}$, large resistivity $\rho_{xx}$) and a small diffusion coefficient in the orthogonal ($\hat x$) direction (small $\sigma_{xx}$, small $\rho_{yy}$).
As a result ~\cite{sammon:2019}, if $N$ is not too large
\begin{equation}\label{eq:ratio}
	\frac{\rho_{xx}}{\rho_{yy}} \simeq \qty(\frac{ \tilde\sigma_0}{8\gamma \alpha^2 N^2})^2 \gg 1,
\end{equation}
where $\tilde\sigma_0 = n_e h \tau / m^{\star}$ is the Drude conductivity at $B=0$ in units of $e^2/h$,  $n_e$ is the electron density, $\tau$ is the momentum relaxation time,
$m^{\star}$ is the electron effective mass, and $\gamma$ is a discussed below numerical factor depending on the nature of scattering.
To derive \req{eq:ratio} we used the HF potential, shown in \rfig{fig:gamma_s}, with the amplitude $\Gamma_s \simeq 0.43 \hbar \omega_c/\alpha$~\cite{fogler:1996}, where $\omega_c$ is the cyclotron frequency, and $\alpha \simeq 18$ is the ratio of the density of states (DOS) $g_B$ in the middle of a spin split LL to that without magnetic field, but per spin, $g_0$~\cite{sammon:2019}. 
In \rref{sammon:2019} we showed that \req{eq:ratio} agrees well with the data from high mobility samples.

\begin{figure}[t]
	\centering
	\includegraphics[width=\linewidth]{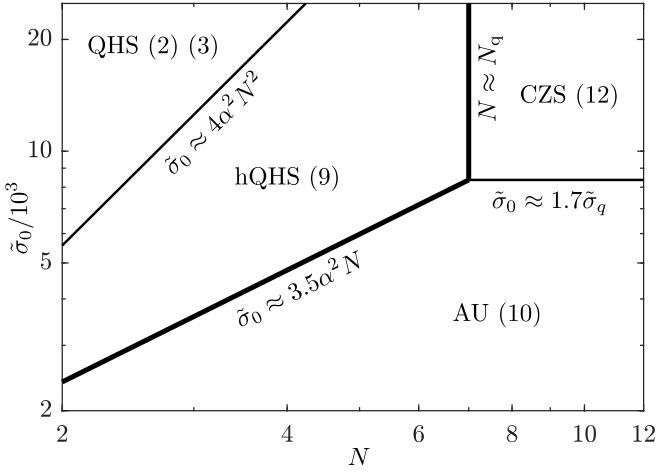}
	\vspace{-0.2 in}
	\caption{Phase diagram 
	for $\tilde{\rho}$
	in the $(N, \tilde\sigma_0)$-plane.
	In the QHS phase $\tilde{\rho}_{xx} \gg \tilde{\rho}_{yy}$, while in the hQHS, AU, and CZS phases $\tilde{\rho}_{xx} = \tilde{\rho}_{yy}$.
	Numbers in parentheses label equations for $\tilde\rho$ in corresponding phases.
	Thick boundaries mark destruction of stripe phases where $\Gamma_i \sim \Gamma_s$.
	$N_{\rm q}$ and $\tilde\sigma_{\rm q}$ are given by Eqs.~\eqref{eq:border_34} and \eqref{eq:border_124}, respectively.
	For GaAs samples with $n_e = 3 \times 10^{11}$ cm$^{-2}$, we used $\tq = 150$ps.}
	\label{fig:phase}

\end{figure}

At large enough $N$, \req{eq:ratio} predicts that the anisotropy of resistivity vanishes. 
In this Letter we theoretically study $\rho(N, \tilde\sigma_0)$ at half-integer $\nu$ in emerging at such $N$ isotropic phase. 
Our results are summarized in the ``phase diagram'' of $\tilde{\rho} \equiv (e^2/h) \rho(N, \tilde \sigma_0)$ depicted in \rfig{fig:phase}. In the
top-left corner it shows the anisotropic QHS phase, discussed above. 
The remaining three phases are isotropic. The Ando and Unemura (AU) phase~\citep{ando:1974.a}, as well as the Coleridge, Zawadzki, and Sachrajda (CZS) phase~\cite{coleridge:1994}
correspond to a regime in which the LL width due to impurity scattering $\Gamma_i$ dwarfs the amplitude of the HF potential of stripes $\Gamma_s$ so that the stripes are destroyed by disorder.
As a result, in both phases $\tilde{\rho} \propto 1/N$.
However, $\tilde{\rho}$ of the two phases differ by the ratio $\tau/\tq$ of momentum and quantum relaxation times. 
Indeed, the AU phase corresponds to low-mobility samples in which the short range scattering determines both scattering times, and $\tau/\tq = 1$, while the CZS phase corresponds to the high mobility samples where scattering on Coulomb impurities leads to  $\tau/\tq \gg 1$.

\begin{figure}[t]
    \centering
    \includegraphics[width = 0.85\linewidth]{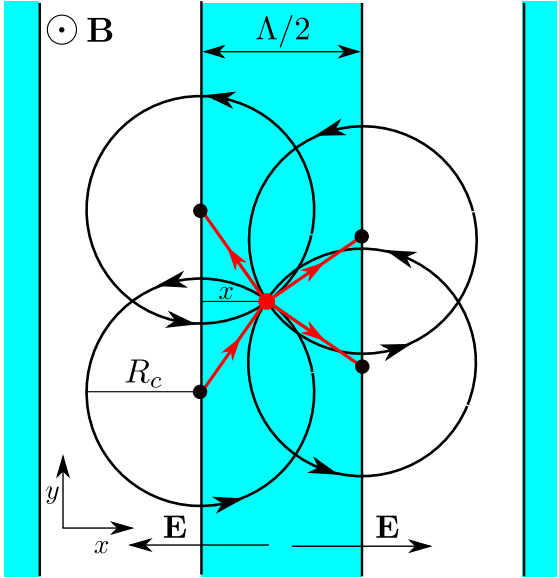}
    \caption{Impurity scattering dominated hopping transport in hidden QHS phase in the quasi-classical (large $N$) limit. An electron with the guiding center (black dot) at the lower left edge of the central electron stripe is scattered off an impurity (red dot) at the distance $x$ from its edge. Three possible hops of the guiding center are shown by red arrows.} 
    \label{fig:typical_scat}
\end{figure}
On the other hand, to the best of our knowledge, the third isotropic phase, located between the QHS phase on one side and the AU/CZS phases has not been discussed in the literature. In this phase $\Gamma_s \gg \Gamma_i $ so that electrons still form stripes, but the stripes do not lead to strong anisotropy of resistivity, because the drift of cyclotron center along $y$ direction produces smaller contribution to conductivity than impurity scattering, which leads to hops of the cyclotron center in all directions at the distance of the order of $R_c$ (see \rfig{fig:typical_scat}). Although, generally speaking this is not enough to make conductivity of an anisotropic system isotropic we will show below that for QHS with period $\Lambda=2.84 R_c$ and large $N$ resistivity anisotropy does not exceed two percents. Therefore, in a semi-quantitative theory, we treat this phase as an isotropic one and call it the ``hidden QHS" (hQHS) phase. 
As in the QHS phase, in the hQHS phase stripes still determine the density of states. We will see that as a result, in the hQHS phase $\tilde{\rho}(N,\tilde{\sigma}_0)$ is independent of $N$. 

Let us now derive the borders of all four phases and the expressions for $\tilde{\rho}(N,\tilde{\sigma}_0)$ for a series of samples with approximately the same $n_e$ and widely varying mobility, which are made of very high mobility GaAs quantum wells by replacing small and varying fraction $x$ of Ga atoms by Al~\cite{gardner:2013}. 
In these samples, the short range Al impurities determine momentum relaxation times $\tau$ and $\tau_B$ at $B=0$ and strong $B$ correspondingly, while $\tq$ is determined by scattering on Coulomb background impurities and remote donors and, therefore, is independent on $x$~\cite{sammon:2018}. We show below that for such samples $\gamma \simeq 0.5$.

\emph{QHS phase}. 
Combining \req{eq:ratio} with $\gamma \simeq 0.5$ and Eq.\,(36) in \rref{macdonald:2000}, $(\tilde\rho_{xx} \tilde \rho_{yy})^{1/2} \simeq 1/8 N^{2}$, we find
\begin{equation}
	\tilde\rho_{xx} \simeq \frac{\tilde\sigma_0}{ 32 \alpha^2 N^4}\,,
\label{eq:rho_xx}
\end{equation} 
\begin{equation}
	\tilde\rho_{yy} \simeq \frac{\alpha^2 }{2 \tilde\sigma_0}\,. \label{eq:rho_yy}
\end{equation}
For a given $\tilde\sigma_0$ the ``hard'' $\tilde\rho_{xx}$ scales with $N^{-4}$ whereas the ``easy'' $\tilde \rho_{yy}$ is $N$-independent. 
The border between the QHS and hQHS phases in \rfig{fig:phase} is determined by the condition $\tilde\rho_{xx} \approx \tilde\rho_{yy}$ or
\begin{equation}\label{eq:border_12}
	\tilde\sigma_0 \approx 4 \alpha^2 N^2\,,
\end{equation} 

\emph{hQHS phase}.
We show below that the hQHS phase resides between its upper border, 
Eq.~\eqref{eq:border_12}, and its lower border 
\begin{equation}\label{eq:border_23}
    \tilde\sigma_0 \approx 3.5 \alpha^2 N\,.
\end{equation}
To find $\tilde\rho(N, \tilde\sigma_0)$, we start with Eqs.~(38) and (39) in Ref.~\onlinecite{dmitriev:2012} 
\begin{equation}\label{eq:39}
	\tilde\sigma =  \frac{h v_{F}^{2} g_{B}\tau_{B}}{2(1+ \omega_c^2 \tau_B^2)} = \frac{\tilde\sigma_0}{2 (1+ \omega_c^2 \tau_B^2)}\,,
\end{equation}
where
\begin{equation}\label{eq:tau_B}
	\frac{1}{\tau_B} \simeq \frac{1}{\tau} \frac{g_B}{g_0}\,.
\end{equation}
Here, $\tau_B$ and $g_B$ are the scattering time and the DOS at the center of the Landau level in strong magnetic field $B$, while $v_F$ and $g_0 = m^{\star}/2\pi \hbar^2$ are the Fermi velocity and the DOS per spin at $B = 0$. 
Our \req{eq:39} has a factor $1/2$ compared to Eq.\,(39) of \rref{dmitriev:2012} because we deal with spin resolved LLs. 
\footnote{Note that $\tau_B$ used in this paper is four times smaller than one used in \rref{sammon:2019}}.

In the hQHS phase, $g_B = \alpha g_0$~\cite{sammon:2019}, and due to Eq.~\eqref{eq:border_23} we have the double inequality $\tilde\sigma_0 \gg \alpha^2 N \gg \alpha N$.
Therefore $\omega_c\tau_B = \tilde{\sigma}_0/2 \alpha N \gg 1$, and \req{eq:39} yields
\begin{equation}\label{eq:sig_xx}
	\tilde\sigma \simeq \frac{2 \alpha^2 N^2}{\tilde\sigma_0 }\,.
\end{equation}
Also the same double inequality implies $\tilde\sigma \ll \tilde\sigma_H \simeq 2N$, and we arrive at
\begin{equation}\label{eq:rho_2}
	\tilde \rho \simeq \frac{\tilde\sigma}{\tilde\sigma_H^2} \simeq \frac{\alpha^2}{2 \tilde\sigma_0},
\end{equation}
which is the same as \req{eq:rho_yy}. The independence of $\tilde \rho$ on $N$ and its inverse proportionality to $\tilde\sigma_0$ are the hallmarks of the hQHS phase. 
In the second part of our paper this result is confirmed by a similar to Ref.~\onlinecite{aizin1984} Kubo formula based calculation of impurity scattering dominated  $\sigma_{xx}$ and $\sigma_{yy}$.  

\emph{AU phase.} Using $\tilde\sigma =  2N/\pi$ calculated in \rref{ando:1974.a} for low mobility samples with $\tau = \tq $ and $\tilde\sigma_H \simeq 2N$ we find
\begin{equation}\label{eq:rho_3}
	\tilde\rho = \frac{\tilde\sigma}{\tilde\sigma^2 + \tilde\sigma_H^2} \simeq \frac{0.14}{N}\,.
\end{equation}
This parameter-free result matches Eq.~\eqref{eq:rho_2} at the upper border of AU phase given by Eq.~\eqref{eq:border_23} and shown in Figure~\ref{fig:phase}. This border can be also obtained by equating $\Gamma_s$ and $\Gamma_i$.

\emph{CZS phase.} To find  $\tilde\rho$ in the CZS phase in samples with $\tau \gg \tq$, we calculate $\tau_B$ using Eq.~\eqref{eq:39} with $g_B = g_0 \sqrt{\omega_c \tq}$ \cite{coleridge:1994,raikh:1993,mirlin:1996}. 
Combining this with $\omega_c \tau_B \sim \sqrt{\omega_c \tau^2/\tq} \gg 1$ \req{eq:39} gives \cite{dmitriev:2012}
\begin{equation}\label{eq:sigma_4}
	\tilde\sigma \simeq \frac{\tq}{\tau} N\,,
\end{equation}
which has an extra factor of $\pi \tq/ 2 \tau$ compared to $\tilde\sigma =  2N/\pi$ in the AU phase \citep{ando:1974.a}.
For $\tq / \tau \ll 1$, we have $\tilde\sigma \ll \tilde\sigma_H$ and 
\begin{equation}\label{eq:rho_4}
	\tilde\rho \simeq \frac {\tilde \sigma} {\tilde \sigma_H^2} = \frac{1}{4} \frac{\tq}{\tau} \frac{1}{N}\,.
\end{equation}
This agrees with Eq.~(6) in Ref.~\onlinecite{coleridge:1994}.
Equation~\eqref{eq:rho_4} matches $\tilde \rho$ in the AU phase, \req{eq:rho_3}, at $\tau \approx 1.7 \tq$ or at
\begin{equation}\label{eq:border_34}
    \tilde\sigma_0 \approx 1.7 \tilde\sigma_{\rm q}\,,~~ \tilde\sigma_{\rm q} \equiv \frac{h n_e \tq}{m^{\star}}\,.
\end{equation} 
Eq.~\eqref{eq:rho_4} also matches $\tilde \rho$ in the hQHS phase, \req{eq:rho_2}, at 
\begin{equation}\label{eq:border_124}
    N \approx N_{\rm q} \equiv \frac{h n_e \tq}{2 \alpha^2 m^{\star}}\,.
\end{equation}
This equation allows one to find $\tq$ using experimental $N_{\rm q}$. 

To construct the phase diagram \rfig{fig:phase}, we used Eqs.~\eqref{eq:border_12}, \eqref{eq:border_23}, \eqref{eq:border_34}, and \eqref{eq:border_124} with $n_e = 3 \times 10^{11}$ cm$^{-2}$, $m^{\star} = 0.067 m_e$, and $\tq = 150$ ps (twice larger than the one in Ref.~~\onlinecite{shi:2017a}). 

Let us now discuss predictions of our phase diagram \rfig{fig:phase} for $\tilde \rho(N) $ of three hypothetical samples with $\tilde\sigma_0 = 2\times 10^3, 5\times 10^3$ and $10^4$. 
The first sample at all $N$ resides in the AU phase and, therefore, should obey Eq.~\eqref{eq:rho_3}. 
The second one at $N < 5$ is in the hQHS phase and, therefore, its $\tilde \rho(N)$ should be given by Eq.~\eqref{eq:rho_2} and be independent on $N$. 
This plateau should end at $N \geq 5$, where $\tilde \rho(N)$ should start declining as $1/N$ according to Eq.~\eqref{eq:rho_3}. 
Finally, the third sample at $N=2$ should show anisotropic $\tilde \rho(N)$, then between $N=3$ and $N=N_{\rm q}=7$, should show a plateau $\tilde \rho(N)$, and eventually at $N > 7$ should follow Eq.~\eqref{eq:rho_4} of the CZS phase. 
Such a diversity of $\tilde \rho(N) $ dependencies is a consequence of the predicted in this paper hidden stripe phase. 
If such $\tilde \rho(N)$ are observed experimentally in samples, one should be able to find $\tq$ from experimental value of $N_{\rm q}$.

In the rest of the paper we justify our results based on a semi-quantitative isotropic approach of Eq.~\eqref{eq:39}.
Let's start with $\sigma_{xx}$ induced by short range impurity scattering. Using the Kubo formula for a sample with sides $L_x$ and $L_y$ we have
\begin{equation}
    \sigma_{xx} = \frac{\pi \hbar e^2}{L_x L_y} \sum_{i, j} \ev{\abs{\dot{X}_{i j}}^2} \delta(E_F - \epsilon_{i}) \delta(E_F - \epsilon_{j}),
\end{equation}
where $E_F$ is the fermi energy, $i$ and $j$ run over all states with energy $\epsilon_{i}$ and $\epsilon_{j}$, and $U(\vb{r}) = U_0 a^3\sum_l \delta^{(3)}(\vb{r} - \vb{r}_l)$ is the impurity potential with range of the lattice constant $a$.
In Landau gauge a wavefunction is given by
\begin{gather}
    \psi_{i}(\vb{r}) = \phi(z) \exp(\frac{-i y X_{i}}{l_B^2})\frac{\chi_N(x-X_{i})}{\sqrt{L_y}}\,,\\
    \chi_N(x) = \frac{\exp(-x^2/2l_B^2) H_N(x/l_B)}{\pi^{1/4} \sqrt{2^N N! l_B}}\,,\\
    \phi(z) = (2/w)^{1/2} \sin(\pi z/w),
\end{gather}
where $w$ is the width of the quantum well.
Using $X = -l_B^2 p_y/\hbar$, the matrix element of the velocity can be written as 
\begin{equation}
    \dot{X}_{i j} = \frac{i}{\hbar} (X_{j} - X_{i}) U_{i j}.
\end{equation}
Then the conductivity becomes 
\begin{equation}
\begin{split}
    \sigma_{xx} =& \frac{\pi e^2}{\hbar L_x L_y} \sum_{i, j} \ev{\abs{U_{i j}}^2} (X_{i} - X_{j})^2 \\
    &\times \delta(E_F - \epsilon_{i}) \delta(E_F - \epsilon_{j}).
\end{split}
\end{equation}
For short range impurities of three-dimensional concentration $N_3$ with the correlator $\ev{U(\vb{r})U(\vb{r}')} = N_3 (U_0 a^3)^2 \delta^{(3)}(\vb{r} - \vb{r}')$, we have
\begin{equation}
    \begin{split}
        \ev{\abs{U_{i j}}^2} &= \int \dd[3]{r} \dd[3]{r'} \psi_{i}^*(\vb{r}) \psi_{j}(\vb{r}) \\
        & \times \psi_{i}(\vb{r'}) \psi_{j}^*(\vb{r'}) \ev{U(\vb{r}) U(\vb{r'})} \\
        &=  \frac{3N_3(U_0 a^3)^2}{2w L_y} \int \dd{x} \chi_N^2(x - X_{i}) \chi_N^2(x - X_{j}).
    \end{split}
\end{equation}
Using $\delta(E_f - \epsilon_{i}) = \sum_m \delta(X_{i} - x^{(m)})/eE$ where $x^{(m)}$ is the $m$th solution of $\epsilon(x) = V(x) = E_F$, $eE = \abs{\dv*{\epsilon}{x}}_{x=x^{(m)}}$, and $\sum_{i} = (L_y/2\pi l_B^2) \int \dd{X_{i}}$, we arrive at 

\begin{equation}\label{eq:kubo_sig_xx}
    \begin{split}
        \sigma_{xx} &= \frac{\pi e^2}{\hbar L_x} \frac{L_y}{(2\pi l_B^2 eE)^2} \int \dd{X_{i}} \dd{X_{j}} \ev{\abs{U_{i j}}^2} \\
    &\times (X_{i} - X_{j})^2 \sum_{m,n} \delta(X_{i} - x^{(m)}) \delta(X_{j} - x^{(n)}) \\ 
     &=\frac{e^2 g_B^2 R_c^2}{2 g_0 \tau} \eta_x.
    \end{split}
\end{equation}
Here we have ignored all terms in the summation with $|X_{i}-X_{j}|>\Lambda/2$, as these terms are exponentially suppressed by the overlap of the wave functions in $U_{i j}$. 
Additionally, we have introduced the transport relaxation rate in zero magnetic field 
\begin{equation}
        \frac{1}{\tau} = \frac{2\pi}{\hbar} g_0 \frac{3N_3(U_0a^3)^2}{2w},
\end{equation}
as well as the dimensionless coefficient
\begin{equation}
    \eta_x = \qty(\frac{\Lambda}{2R_c})^2 \Lambda \int \dd{x} \chi_N^2(x) \chi_N^2(x-\Lambda/2).
\end{equation}
For $\Lambda = 2.84 R_c$ and $N > 2$, $\eta_x(N)=1.07 \pm 0.15$. It oscillates with $N$ and tends to 1.07 at $N \to \infty$.

Now we can show that the coefficient defined in \rref{sammon:2019}  $\gamma \simeq \eta_x/2 \simeq 0.53$ for short range impurities , which was used in Eqs.~\eqref{eq:ratio},~\eqref{eq:rho_xx},~\eqref{eq:rho_yy},~\eqref{eq:border_12}. To this end we should equate the ratio of $\sigma_{xx}/g_{B}e^{2}$ obtained from Eq.~\eqref{eq:kubo_sig_xx} to the expression for $D_{xx}$ obtained from the combination of Eqs. (2) and (7) of \rref{sammon:2019} .

Next we calculate $\sigma_{yy}$ induced by short range impurity scattering by the Kubo formula:
\begin{equation}
    \sigma_{yy} = \frac{\pi \hbar e^2}{L_x L_y} \sum_{i, j} \ev{\abs{\dot{Y}_{i j}}^2} \delta(E_F - \epsilon_{i}) \delta(E_F - \epsilon_{j}).
\end{equation}
The velocity operator along $y$ can be written as 
\begin{equation}
    \dot{Y} = \frac{i}{\hbar} [H,Y] = -\frac{l_B^2}{\hbar} \pdv{U}{x},
\end{equation}
where we ignore the drift in internal electric field $\vb{E}$ of QHS and use $Y = l_B^2 p_x/\hbar$. 
Therefore the matrix element $\dot{Y}_{i j}$ can be evaluated doing integration by parts
\begin{equation}
    \begin{split}
        \dot{Y}_{i j} &= -\frac{l_B^2}{\hbar} \int \dd[3]{r} \pdv{U}{x} \psi_{i}^*(\vb{r}) \psi_{j}(\vb{r}) \\
        &= \frac{l_B^2}{\hbar} \int \dd[3]{r} U(\vb{r}) \pdv{x} \qty[\psi_{i}^*(\vb{r}) \psi_{j}(\vb{r})].
    \end{split}
\end{equation}
After averaging over impurities positions, we get
\begin{equation}
    \begin{split}
        \ev{\abs{\dot{Y}_{i j}}^2} &= \frac{l_B^4}{\hbar^2} \frac{3N_3(U_0 a^3)^2}{2w L_y} \\
        &\times \int \dd{x} \qty{\dv{x}\qty[\chi_N(x - X_{i}) \chi_N(x - X_{j})]}^2.
    \end{split}
\end{equation}
Hence, for the conductivity we get 
\begin{equation}\label{eq:kubo_sig_yy}
    \begin{split}
        \sigma_{yy} &= \frac{\pi e^2}{\hbar L_x} \frac{l_B^4}{(2\pi l_B^2 eE)^2} \int \dd{X_{i}} \dd{X_{j}} \frac{3N_3(U_0a^3)^2}{2w}  \\
    &\times \int \dd{x} \qty{\dv{x}\qty[\chi_N(x - X_{i}) \chi_N(x - X_{j})]}^2  \\
    &\times \sum_{m,n} \delta(X_{i} - x^{(m)}) \delta(X_{j} - x^{(n)}) = \frac{e^2 g_B^2 R_c^2}{2 g_0 \tau} \eta_y,
    \end{split}
\end{equation}
where the summation of the product of delta functions can be evaluated separately. If $m=n$, then there are $2L_x/\Lambda$ terms with $X_{i} = X_{j}$. On the other hand, if $m\neq n$, there are $4L_x/\Lambda$ terms with $\abs{X_{i} - X_{j}} = \Lambda/2$ (all other terms are negligible).
This leads to the coefficient $\eta_y$ in Eq.~\eqref{eq:kubo_sig_yy}
\begin{equation}
\begin{split}
        \eta_y &= \frac{\Lambda l_B^4}{2R_c^2} \int \dd{x} \left\{\qty[\dv{x}\qty(\chi_N^2(x))]^2\right. \\
        &+ \left.2\qty[\dv{x}\qty(\chi_N(x) \chi_N(x - \Lambda/2))]^2\right\}.
    \end{split}
\end{equation}
For $\Lambda = 2.84 R_c$ and $N > 2$, $\eta_y(N) = 1.06 \pm 0.01$. It oscillates with $N$ and tends to 1.06 at $N \to \infty$.

Next we calculate the dimensionless conductivities from Eq.~\eqref{eq:kubo_sig_xx} and \eqref{eq:kubo_sig_yy} using $\tilde{\sigma} = (h/e^2) \sigma$, and arriving at 
\begin{equation}\label{eq:kubo_result}
    \tilde{\sigma}_{xx} = \frac{2\eta_x\alpha^2 N^2}{\tilde{\sigma}_0}\qc \tilde{\sigma}_{yy} = \frac{2\eta_y\alpha^2 N^2}{\tilde{\sigma}_0}.
\end{equation}
Thus, we see that in the hQHS phase the disorder dominated conductivity is isotropic and agrees with Eq.~\eqref{eq:sig_xx} within 15\%. 
This justifies the above use of semi-quantitative isotropic approach based on Eq.~\eqref{eq:39}. 
We would like to emphasize that such a small anisotropy of the conductivity of the hQHS phase is related to the value of the period $\Lambda =2.84 R_c$. 
We checked that at $N \to \infty$ with varying $\Lambda$, the anisotropy vanishes at $\Lambda = 2.82 R_c$.
Therefore, the anisotropy is very small at $\Lambda = 2.84 R_c$. 

We thank I. Dmitriev, M. Fogler, and X. Fu for valuable discussions. Calculations by Y. H. and M. S. were supported primarily by the NSF through the University of Minnesota MRSEC under Award No. DMR-1420013.
M. Z. acknowledges support from the U.S. Department of Energy, Office of Science, Basic Energy Sciences, under Award No. DE-SC0002567.

\medskip
%

\end{document}